\documentclass[twocolumn]{aastex701}
\usepackage{booktabs}

\begin{document}

\title{Environmental Dependence of Galaxy properties: A study of 341 Ring Galaxies in Cosmic Voids}

\author[orcid=0009-0009-6823-8756]{Santosh Poudel}

\affiliation{Central Department of Physics, Tribhuvan University}
\email[show]{santosh.765511@cdp.tu.edu.np}  

\author{Binil Aryal} 

\affiliation{Central Department of Physics, Tribhuvan University}
\email[show]{binil.aryal@cdp.tu.edu.np}

\begin{abstract}

We investigate the morphological and physical properties of ring galaxies residing within cosmic voids. Using void catalogs from VoidFinder, ring candidates identified via the Galaxy Zoo 2 decision tree, and morphological classifications from the Buta (2017) CVRHS based catalog, we analyze a sample of 341 void ring galaxies and  find a radial preference, with 91.5\% located away from the void cores. Morphologically, inner rings and inner pseudorings account for 45.2\% of the sample while outer pseudorings are even more common(56.3\%) and outer rings appear in 17.9\% of galaxies, which points to secular evolution driven by internal dynamics as the primary formation mechanism. Compared to the general void population, our sample ring galaxies are found to be more massive, redder, and have lower specific star formation rates. Subtle gradients in stellar mass and sSFR from void centers to edges also shows a gentle, density-dependent evolutionary progression. Our results show ring galaxies as a distinct, secularly evolved population shaped by the weakest large-scale environmental gradients.

\end{abstract}


\section{Introduction} 

Cosmic voids are the vast underdense regions that make up the cosmic web along with filaments, sheets, tendrils and clusters\citep{montero2025galaxy,argudo2024morphologies,1991ARA&A..29..499G}. Compared to filaments and clusters, they contain relatively much lower number of galaxies\citep{pub.1183713275}, so can be regarded as almost isolated\citep{1755b2c3495c4d95a95cee47eba0a0d0,goldberg2004simulating}. Because of this isolation the galaxies residing inside the cosmic voids evolve differently, resulting in different physical properties as well as morphology. On observation of the void galaxies from different void finding algorithms, void galaxies were found to be more bluer, fainter and having high specific star formation rates compared to the galaxies in denser regions\citep{conrado2024cavity}. These observations were also found to match in simulations\citep{curtis2024properties,rodriguez2024evolutionary}.

In general, the true investigation of evolution of any galaxies requires disentangaling of their external and internal drivers\citep{balogh2002galaxy}. Galaxy- galaxy interaction and mergers are know to have a effect on morphology of the galalxies\citep{kaviraj2025mergers,lambas2012galaxy}. Therfore studying galaxies residing in an isolated enviorments provides us a baseline for understanding evolution driven primarily by internal secular processes. 

In comparison to the galaxies in denser regions, isolated galaxies were seen to have longer bars, lower assymetry, reduced central concentration and decreased clumpiness\citep{durbala2008photometric}. The high presence of grand-design spirals in nonbarred isolated galaxies shows us that coherent, large-scale spiral arms can form and be maintained without external tidal triggers which points us in the direction to the importance of internal dynamical processes, such as global disk instabilities, in generating these patterns\citep{buta2019comprehensive}. Furthermore, the predominance of pseudobulges over classical bulges seen in these isolated spiral suggests that their central mass concentrations were built gradually through secular processes, such as bar-driven gas inflow, rather than through the rapid, merger-driven assembly thought to form classical bulges\citep{buta2013galaxy}.

Since the study of isolated field galaxies have been invaluable, void galaxies can represent the next frontier in low-density exploration. Among several galactic features rings play a vital role in the life and the lifetime of its host\citep{1979ApJ...227..714K}.

In this work we aim the  following: (a) Whether ring galaxies in voids are predominantly shaped by secular evolution or external interactions, (b) Whether the physical properties of ring galaxies exhibit gradients with void-centric distance and (c) Whether ring galaxies show a radial preference within cosmic voids.

\section{Data} \label{sec:style}

\subsection{Identification of Voids}

We make use of the void catalogs produced by Douglass et al\citep{Douglass_2023}. They produced three separate void catalogs  by the use of two different void finding algorithms one void finder\citep{el1997voids} which employs sphere growing technique and other two namely VIDE\citep{sutter2015vide} and REVOLVER based on ZOBOV\citep{neyrinck2008zobov}. Among the three catalogs we used the void catalog\footnote{\url{https://zenodo.org/records/11043278}} generated via voidfinder because it excludes the shell crossing regions and cleanly isolates the pristine void interior which make them suitable for void galaxy studies. Our catalog had a total of 1163 voids.

VoidFinder takes a volume-limited galaxy catalog as an input and seggregate the galaxies  into wall and field populations based on third-nearest neighbor distances. Galaxies whose third-nearest neighbor exceeds $d_3 + 1.5\sigma d_3$, where $d_3$ is the mean separation and $\sigma d_3$ its standard deviation, are classified as isolated field galaxies and removed from the sample. This leaves behind the clustered wall galaxies that define the boundaries of potential voids.
These wall galaxies are placed onto a three-dimensional grid with cell sizes of $5 h^{-1}$ Mpc. From each empty grid cell, a sphere is grown outward until its surface contacts four wall galaxies. Any sphere extending more than 10\% beyond survey boundaries is discarded to prevent edge effects.
The algorithm then identifies maximal spheres,the largest possible sphere that can fit inside a distinct void region. All spheres exceeding $10 h^{-1}$ Mpc in radius are sorted and evaluated in descending order. A sphere becomes a maximal sphere only if it overlaps no previously identified larger maximal sphere by more than 10\% of its volume, ensuring each void region remains dynamically separate.
Finally, all remaining non-maximal spheres are merged with a single maximal sphere if they overlap it by at least 50\% of their volume. Each void is thus defined as the union of its maximal sphere and these associated smaller spheres who in results captures the irregular geometry of  cosmic voids rather than simple spherical approximations.

\subsection{Galaxy zoo}
Galaxy Zoo is a citizen science project that first originated to classify the morphologies of large galaxies samples obtained from the Sloan Digital Sky Survey (SDSS) at the Apache Point Observatory in New Mexico, USA. The first Galaxy Zoo project \citep{lintott2008galaxy} was made possible by hundreds of thousands of volunteer citizen scientists who visually classified nearly one million galaxies based on basic morphological features namely elliptical, spiral, or merger. This was because expert visual inspection was impractical and may took years because of the large size of the project and automated classification algorithms had limitations at the time.

Subsequently, Galaxy Zoo 2 \citep{willett2013galaxy} was introduced and was completed in 2013 following the success of first galaxy zoo project. This time it used a rigorous decision tree to record finer morphological details such as galactic bars, bulge prominence, spiral arm structure, tightness and multiplicity, edge-on disk properties, and other features such as rings and merger signatures. The GZ2 sample comprises $304{,}122$ galaxies from SDSS Data Release 7 with apparent magnitude $m_r < 17.0$, angular size $r_{90} > 3^{\prime\prime}$, and spectroscopic redshifts in the range $0.0005 < z < 0.25$.

The classification process used a web-based interface that directed volunteers through a decision tree of 11 tasks with 37 possible responses. More than 16 million classifications were collected, with each galaxy receiving a median of 44 independent classifications. To ensure reliability, an iterative weighting scheme down-weighted contributions from classifiers showing poor consistency with the majority consensus, resulting in approximately 95\% of classifiers receiving full weight. The final morphological parameters were corrected for classification bias which would arise because of the systematic tendency for finer features to be underidentified in smaller, fainter, more distant galaxies. These debiased vote fractions provide probabilistic estimates of morphological features and showed excellent agreement (approx.90 \%) with expert classifications.
The decision tree is given in \citep{willett2013galaxy}.

\subsection{Sample Selection}
\label{ssec:sample}
Since the void catalog was constructed from a NASA-Sloan Atlas (NSA) \footnote{\url{https://www.nsatlas.org/}} volume‑limited sample \citep{blanton2005new}, we first identified all galaxies residing inside the defined cosmic voids. Douglass et al.\citep{Douglass_2023} report that 17.6\% of the volume‑limited galaxies lie in voids when considering all 1163 voids. However, the VoidFinder maximal spheres file includes an \texttt{edge\_flag} column indicating whether a void is potentially affected by the survey boundary (flag = 1 if any part of the void extends beyond the mask, flag = 2 if any sphere centre lies within 10\,Mpc/$h$ of the mask). To avoid boundary effects, we restrict our analysis to voids fully contained within the survey volume, i.e., those with \texttt{edge\_flag = 0}. For each such void, given the Cartesian coordinates of its maximal sphere centre and its effective radius, we considered a galaxy to be within void if its position lies within any of the void's constituent spheres. For each void galaxy we then compute the normalized distance to the void centre as $r/R$, where $r$ is the distance from the galaxy to the maximal sphere centre and $R$ is the void's effective radius. We then cross‑matched these void galaxies with the Galaxy Zoo 2 (GZ2) morphological catalog, using a positional tolerance of 1\arcsec\ to ensure reliable counterparts. From the matched sample, we applied the GZ2 decision‑tree logic to isolate ring‑galaxy candidates. Specifically, we required galaxies to have a debiased vote fraction indicative of a definite ring structure.

\begin{figure*}[hbt!]
\centering

\begin{tabular}{cccc}
\includegraphics[width=0.23\textwidth]{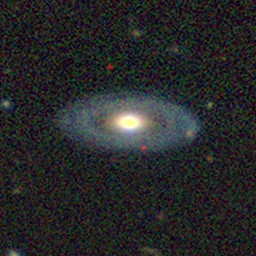} &
\includegraphics[width=0.23\textwidth]{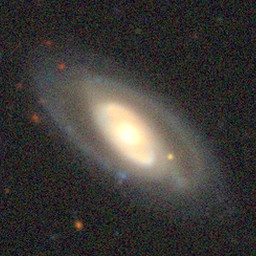} &
\includegraphics[width=0.23\textwidth]{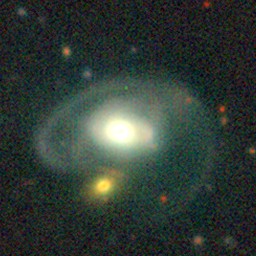} &
\includegraphics[width=0.23\textwidth]{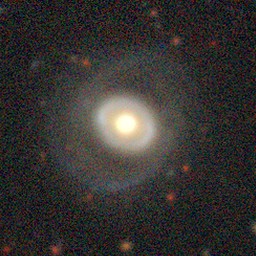} \\

\end{tabular}

\caption{
Images of four ring galaxies residing at outskirts of voids from our sample.
 NSAID 263999 ($z=0.07656$, Void 435 ): An early-type lenticular galaxy (S0\(^{+}\)) featuring an outer ring and an inner lens, with no bar;
 NSAID 238331 ($z=0.032417$, Void 32): An intermediate-type spiral galaxy featuring an outer pseudoring intermediate bar and an inner pseudoring;
 NSAID 273968 ($z=0.057957$, Void 658): An early type spiral galaxy with a weak central bar, a distinct outer pseudoring, and visible morphological peculiarities.
 NSAID 369595 ($z=0.047381$, Void 295): A transition type galaxy without a central bar featuring a well-defined inner ring and a double outer pseudoring.
}
\label{fig:example_galaxies1}
\end{figure*}

Given the subtle and sometimes ambiguous appearance of ring structures in survey images and to avoid contamination by artifacts, overlapping objects, or misclassified features, we performed a final visual inspection of every candidate. Each of the galaxies were observed and inspected individually. This inspection filtered out objects where the ring-like morphology was spurious and we retained only those systems that exhibited a definite, coherent ring or ring-like morphology.

After this multi-stage selection,void identification, Galaxy Zoo cross-matching, ring-candidate filtering, and visual verification, we arrived at a  sample of 341 ring galaxies which formed the basis for all subsequent analysis presented in this work.

\subsection{The Sample of Void-Residing Ring Galaxies, their morphologies and calculation of Physical Properties}
\label{ssec:sample_props}

Our final sample consists of 341 ring galaxies. The galaxies span a redshift range of $0.01066< z < 0.109723$, with a mean redshift of $z = 0.06443$, which placed them in the local universe where detailed morphological classification is most reliable. The distribution of these galaxies within their host voids is described in Section~\ref{ssec:spatial}.

\begin{figure*}[hbt!]
\centering

\begin{tabular}{cccc}
\includegraphics[width=0.23\textwidth]{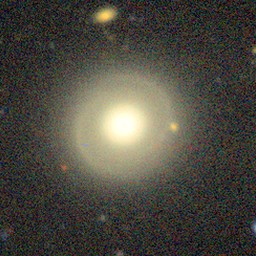} &
\includegraphics[width=0.23\textwidth]{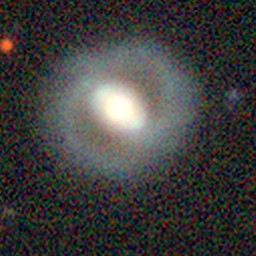} &
\includegraphics[width=0.23\textwidth]{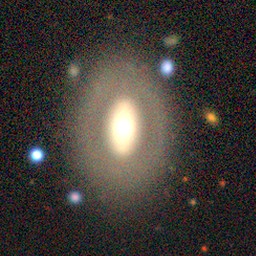} &
\includegraphics[width=0.23\textwidth]{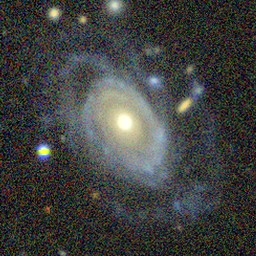} \\

\end{tabular}

\caption{
Images of four ring galaxies residing on the onset of voids edges from our sample.
 NSAID 379096 ($z=0.032593$, Void 689 ): An unbarred, transition type galaxy featuring an inner pseudoring  marking the emergence of spiral structure within a largely smooth, lenticular like disk.;
 NSAID 243707 ($z=0.0341$, Void 95): Inner Pseudoring structure;
 NSAID 56324 ($z=0.042552$, Void 18): An unbarred to weakly barred transition galaxy featuring an outer ring lens, an inner lens, and a late-stage lenticular disk;
 NSAID 501052 ($z=0.077945$, Void 48): Unbarred, intermediate type spiral galaxy featuring an outer pseudoring(uncertain) and a transitional inner ring-spiral structure
}
\label{fig:example_galaxies2}
\end{figure*}

\begin{figure*}[hbt!]
\centering
\begin{tabular}{cccc}
\includegraphics[width=0.23\textwidth]{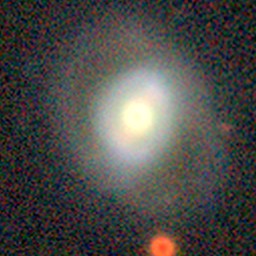} &
\includegraphics[width=0.23\textwidth]{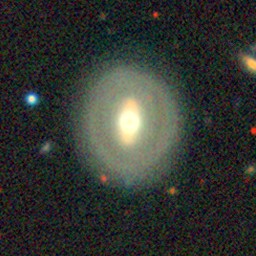} &
\includegraphics[width=0.23\textwidth]{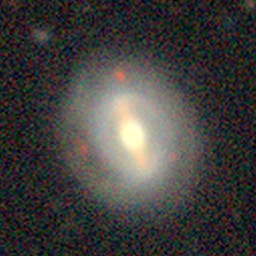} &
\includegraphics[width=0.23\textwidth]{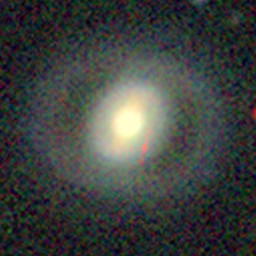} \\

\end{tabular}

\caption{
Images of four ring galaxies residing on the cores of void from our sample.
NSAID 383498 ($z=0.04399$, Void 138 ): An intermediate type spiral galaxy featuring an close outer ring and an intermediate bar ;
 NSAID 530870 ($z=0.0422$, Void 18): An early type spiral lenticular transition galaxy featuring an outer pseudoring, an intermediate bar, and inner spiral arms;
 NSAID 248444 ($z=0.0395$, Void 516): Outer Pseudoring and an starforming inner ring;
 NSAID 449586 ($z=0.0584$, Void 875): An early type spiral lenticular transition galaxy featuring a closed outer ring, an intermediate bar, and a closed inner ring.
}
\label{fig:example_galaxies3}
\end{figure*}

Beyond their environmental properties, to determine the morphological type of each ring structure, we adopted and make use the catalog of \citet{buta2017galactic}. This catalog utilizes the CVRHS (Comprehensive de Vaucouleurs revised Hubble-Sandage) system, which provides standardized classes for ringed galaxies\citep{1959HDP....53..275D}. Our cross-match assigned each of our 341 void galaxies a primary ring type from this system, such as inner ring, inner pseudoring, or ring-lens. The complete distribution of these morphological types within our sample is presented in Table~\ref{tab:morph_summary}. Moreover, the images of twelve ring galaxies residing in three radial bins are shown in Figure~\ref{fig:example_galaxies1},Figure~\ref{fig:example_galaxies2} and Figure~\ref{fig:example_galaxies3} which were taken from the Legacy Surveys Sky Viewer\footnote{\url{https://decaps.legacysurvey.org/viewer}} using DR10 images.

For the calculation of physical properties like stellar mass and color, we use the k corrected  values available in the NSA catalog. We make use of SFR and sSFR values from the MPA-JHU value added galaxy catalog\footnote{\url{https://wwwmpa.mpa-garching.mpg.de/SDSS/DR7/sfrs.html}}, where they were derived by using the techniques based on \citep{2004MNRAS.351.1151B}.

\begin{deluxetable*}{@{\extracolsep{\fill}}lcccc}
\tablewidth{0pt}
\tabletypesize{\small}
\tablecaption{Summary of Morphological Features from the CVRHS Classification System\tablenotemark{a} by Radial Position in Cosmic Voids \label{tab:morph_summary}}

\tablehead{
\colhead{Feature} & 
\colhead{\shortstack{Void Core \\ ($r/R<0.7$)}} & 
\colhead{\shortstack{Edge Onset \\ ($0.7\le r/R<0.95$)}} & 
\colhead{\shortstack{Void Outskirts \\ ($r/R\ge0.95$)}} & 
\colhead{Total} \\
\colhead{} & 
\colhead{($N=29$)} & 
\colhead{($N=211$)} & 
\colhead{($N=101$)} & 
\colhead{($N=341$)}
}
\startdata
\hline
\multicolumn{5}{c}{\textbf{Bar Properties}} \\
SA (non-barred)               & 4 (13.8\%)  & 58 (27.5\%)  & 27 (26.7\%)  & 89 (26.1\%) \\
SAB (weakly barred)\tablenotemark{d} & 24 (82.8\%) & 124 (58.8\%) & 58 (57.4\%)  & 206 (60.4\%) \\
SB (strongly barred)          & 1 (3.4\%)   & 29 (13.7\%)  & 16 (15.8\%)  & 46 (13.5\%) \\
Bars with ansae ($\_a$)       & 4 (13.8\%)  & 33 (15.6\%)  & 22 (21.8\%)  & 59 (17.3\%) \\
\hline
\multicolumn{5}{c}{\textbf{Inner Variety}} \\
(r)  -- inner ring            & 9 (31.0\%)  & 76 (36.0\%)  & 35 (34.7\%)  & 120 (35.2\%) \\
(rs) -- inner pseudoring      & 2 (6.9\%)   & 24 (11.4\%)  & 8 (7.9\%)    & 34 (10.0\%) \\
(s)  -- spiral, no ring       & 3 (10.3\%)  & 17 (8.1\%)   & 10 (9.9\%)   & 30 (8.8\%) \\
(rl) -- inner ring-lens       & 5 (17.2\%)  & 9 (4.3\%)    & 6 (5.9\%)    & 20 (5.9\%) \\
(l)  -- inner lens            & 7 (24.1\%)  & 58 (27.5\%)  & 28 (27.7\%)  & 93 (27.3\%) \\
(bl) -- boxy/peanut inner     & 1 (3.4\%)   & 18 (8.5\%)   & 14 (13.9\%)  & 33 (9.7\%) \\
(p)  -- peculiar inner        & 0 (0.0\%)   & 3 (1.4\%)    & 5 (5.0\%)    & 8 (2.3\%) \\
\hline
\multicolumn{5}{c}{\textbf{Outer Variety}} \\
(R)  -- outer ring            & 3 (10.3\%)  & 42 (19.9\%)  & 16 (15.8\%)  & 61 (17.9\%) \\
(R1) -- outer R1 ring         & 3 (10.3\%)  & 16 (7.6\%)   & 3 (3.0\%)    & 22 (6.5\%) \\
(R1') -- outer pseudoring (R1) & 6 (20.7\%)  & 31 (14.7\%)  & 22 (21.8\%)  & 59 (17.3\%) \\
(R2') -- outer pseudoring (R2) & 4 (13.8\%)  & 26 (12.3\%)  & 13 (12.9\%)  & 43 (12.6\%) \\
(R1R2') -- combined resonance & 4 (13.8\%)  & 11 (5.2\%)   & 10 (9.9\%)   & 25 (7.3\%) \\
(R') -- generic pseudoring    & 19 (65.5\%) & 112 (53.1\%) & 61 (60.4\%)  & 192 (56.3\%) \\
(RL) -- outer ring-lens       & 2 (6.9\%)   & 23 (10.9\%)  & 9 (8.9\%)    & 34 (10.0\%) \\
(L)  -- outer lens            & 2 (6.9\%)   & 6 (2.8\%)    & 2 (2.0\%)    & 10 (2.9\%) \\
Multiple outer features       & 18 (62.1\%) & 92 (43.6\%)  & 48 (47.5\%)  & 158 (46.3\%) \\
\hline
\multicolumn{5}{c}{\textbf{Special Features}} \\
X / boxy/peanut               & 0 (0.0\%)   & 1 (0.5\%)    & 0 (0.0\%)    & 1 (0.3\%) \\
Peculiar (pec)                & 2 (6.9\%)   & 12 (5.7\%)   & 6 (5.9\%)    & 20 (5.9\%) \\
\hline
\multicolumn{5}{c}{\textbf{Classification Quality}} \\
Stage uncertainty ($:$)       & 7 (24.1\%)  & 42 (19.9\%)  & 15 (14.9\%)  & 64 (18.8\%) \\
\hline
\multicolumn{5}{c}{\textbf{Combined Feature Counts}} \\
Any lens feature$^c$          & 9 (31.0\%)  & 64 (30.3\%)  & 30 (29.7\%)  & 103 (30.2\%) \\
\enddata
\tablenotetext{a}{CVRHS: Comprehensive de Vaucouleurs Revised Hubble-Sandage classification system, as described in Buta (2017). The classifications are taken from the catalog of \citep{buta2017galactic} and provide detailed morphological types including bar strength, inner and outer ring structures, lenses, and nuclear features.}
\tablenotetext{b}{OLR subclasses include (R1), (R1'), (R2'), (R1R2') – outer rings/pseudorings associated with the outer Lindblad resonance. Counts are the number of galaxies exhibiting any of these features.}
\tablenotetext{c}{Any lens includes (l), (rl), (L), (RL), (bl). Nuclear features are extremely rare ($<$2\%) and omitted for brevity.}
\tablenotetext{d}{SAB includes all weak/intermediate bar notations (S\_AB, SA\_B, etc.) }
\tablecomments{
Percentages are relative to the total number of galaxies in each radial bin ($N$). For a complete description of the CVRHS notation and feature definitions, we refer the reader to Buta (2017). The table includes only features with non‑negligible counts or clear environmental trends; the full dataset is available in the electronic edition.
}
\end{deluxetable*}

\begin{deluxetable*}{lcccc}
\tabletypesize{\scriptsize}
\tablecaption{Physical Properties of Void Ring Galaxies by Radial Position \label{tab:physical_properties}}
\tablewidth{\textwidth}
\tablehead{
\colhead{Property} & 
\colhead{\shortstack{Void Core \\ ($r/R<0.7$)}} & 
\colhead{\shortstack{Edge Onset \\ ($0.7\leq r/R<0.95$)}} & 
\colhead{\shortstack{Void Outskirts \\ ($r/R\geq0.95$)}} &
\colhead{General Void Galaxies} \\
\colhead{} & 
\colhead{$N=29$} & 
\colhead{$N=211$} & 
\colhead{$N=101$} &
\colhead{\cite{pub.1183713275}}
}
\startdata
\multicolumn{5}{c}{\textbf{Stellar Mass Properties}} \\
\midrule
Mean $\log(M_*/M_\odot)$ & $10.330 \pm 0.180$ & $10.39 \pm 0.180$ & $10.41 \pm 0.18$ & $9.786 \pm 0.002$ \\
Median $\log(M_*/M_\odot)$ & $10.310$ & $10.380$ & $10.390$ & $9.910$ \\
Range $\log(M_*/M_\odot)$ & $10.040$--$10.840$ & $9.860$--$10.950$ & $9.950$--$10.870$ & --- \\
\midrule
\multicolumn{5}{c}{\textbf{Optical Colors (mag)}} \\
\midrule
Mean $u-r$ & $2.230 \pm 0.376$ & $2.255 \pm 0.346$ & $2.238 \pm 0.279$ & $1.763 \pm 0.002$ \\
Median $u-r$ & $2.237$ & $2.262$ & $2.230$ & $1.73$ \\
Blue Fraction ($u-r<2.3$)\tablenotemark{a} (\%)& $55.2 \pm 9.2$ & $54.0 \pm 3.4$ & $57.4 \pm 4.9$ & --- \\
\midrule
Mean $g-r$ & $0.722 \pm 0.088$ & $0.744 \pm 0.098$ & $0.738 \pm 0.098$ & $0.561 \pm 0.0004$ \\
Median $g-r$ & $0.731$ & $0.761$ & $0.760$ & $0.57$ \\
Blue Fraction ($g-r<0.7$)\tablenotemark{b} (\%) & $34.5 \pm 8.8$ & $28.4 \pm 3.1$ & $31.7 \pm 4.6$ & --- \\
\midrule
\multicolumn{5}{c}{\textbf{Star Formation Activity}} \\
\midrule
Mean $\log(\mathrm{SFR})$ [$\log(M_\odot\,\mathrm{yr}^{-1})$] & $-0.346 \pm 0.607$ & $-0.454 \pm 0.632$ & $-0.402 \pm 0.667$ & $-0.30 \pm 0.002$ \\
Median $\log(\mathrm{SFR})$ [$\log(M_\odot\,\mathrm{yr}^{-1})$] & $-0.462$ & $-0.582$ & $-0.598$ & $-0.22$ \\
\midrule
Mean $\log(\mathrm{sSFR})$ [$\log(\mathrm{yr}^{-1})$] & $-11.104 \pm 0.700$ & $-11.263 \pm 0.683$ & $-11.247 \pm 0.715$ & $-10.49 \pm 0.003$ \\
Median $\log(\mathrm{sSFR})$ [$\log(\mathrm{yr}^{-1})$] & $-11.108$ & $-11.451$ & $-11.502$ & $-10.23$ \\
\midrule
\multicolumn{5}{c}{\textbf{Radial Distribution within Voids}} \\
\midrule
Galaxy Fraction (\%) & $8.5$ & $61.9$ & $29.6$ & --- \\
\enddata

\tablenotetext{a}{Blue galaxies defined as $u-r < 2.3$ mag (\citep{strateva2001color})}
\tablenotetext{b}{Blue galaxies defined as $g-r < 0.7$ mag}
\tablecomments{Comparison with general void galaxy population from \cite{pub.1183713275}: Ring galaxies are more massive ($\Delta\log M_* \approx +0.6$ dex), redder ($\Delta(u-r) \approx +0.5$ mag), and have lower sSFRs ($\Delta\log\mathrm{sSFR} \approx -0.8$ dex) than average void galaxies.}
\end{deluxetable*}

\section{Results}
To investigate how galaxy properties vary with distance from the void centre, we divide the sample into three radial bins based on $r/R$. Then, the  choice of the bin boundaries were guided by two considerations: (i) the need to have statistically meaningful numbers in each bin, and (ii) the need to sample the inner, intermediate, and outer parts of the voids separately. Because ring galaxies are strongly concentrated toward the edge, we set the inner boundary at $r/R = 0.7$, which isolates the sparsely populated core (29 galaxies, 8.5\% of the sample). The outer region is further split at $r/R = 0.95$ to distinguish the extreme outskirts from the main edge population. This seggregation gave three bins as void core ($r/R < 0.7$, $N = 29$), edge onset ($0.7 \le r/R < 0.95$, $N = 211$), and void outskirts ($r/R \ge 0.95$, $N = 101$). 
\subsection{Distribution within Voids}
\label{ssec:spatial}

\begin{figure*}[hbt!]
\centering
\begin{tabular}{ccc}
\includegraphics[width=0.31\textwidth]{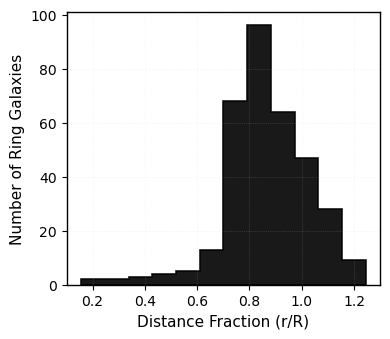} &
\includegraphics[width=0.31\textwidth]{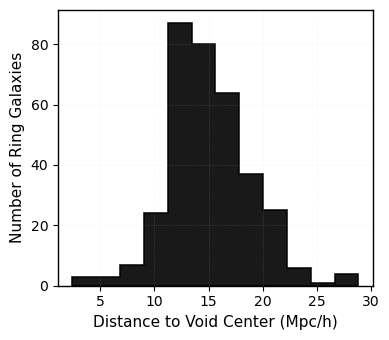} &
\includegraphics[width=0.31\textwidth]{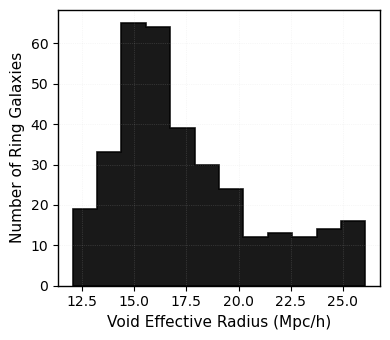} \\

\end{tabular}
\caption{Environmental properties of ring galaxies in cosmic voids. Left: Distribution of distance fraction $(r/R)$, where r is the distace from galaxy to the void center and R is the void effective radius, here the values near zero corresponds to the galaxies near void centers and values around one corresponds to galaxies near void edges. Middle: Distribution of absolute distances to void centers in Mpc/h. Right: Distribution of void effective radii containing ring galaxies. We see majority of ring galaxies are in medium sized voids.}
\label{fig:environment}
\end{figure*}

The distribution of ring galaxies within their host voids is shown in Figure~\ref{fig:environment}. The distribution spans the range from the deep void interior ($r/R \sim 0.156$) to the edges ($r/R = 1.245$), but they are heavily concentrated towards the outer regions. Only 8.5\% of the sample resides in the void core, while 61.9\% lie in the edge‑onset region and 29.6\% in the outermost outskirts.
Moreover, the distribution of effective radii of the voids hosting ring galaxies shows that ring galaxies are found in voids of various sizes, from moderate ($12$ Mpc$h^{-1}$) to large ($27$ Mpc$h^{-1}$).

\subsection{Stellar Mass Distribution}
\label{ssec:stellar_mass}

\begin{figure*}[hbt!]
\centering
\begin{tabular}{ccc}
\includegraphics[width=0.31\textwidth]{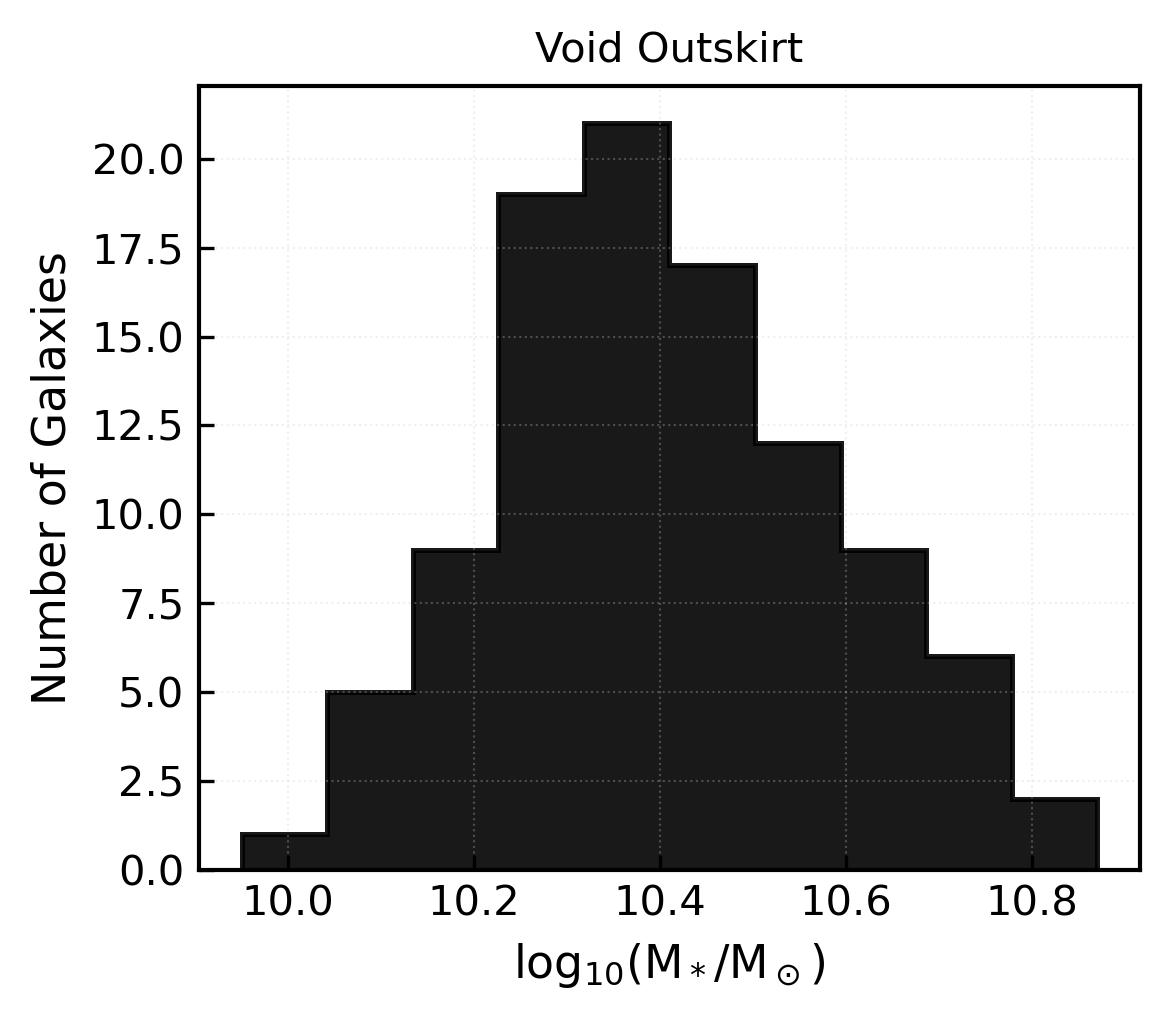} &
\includegraphics[width=0.31\textwidth]{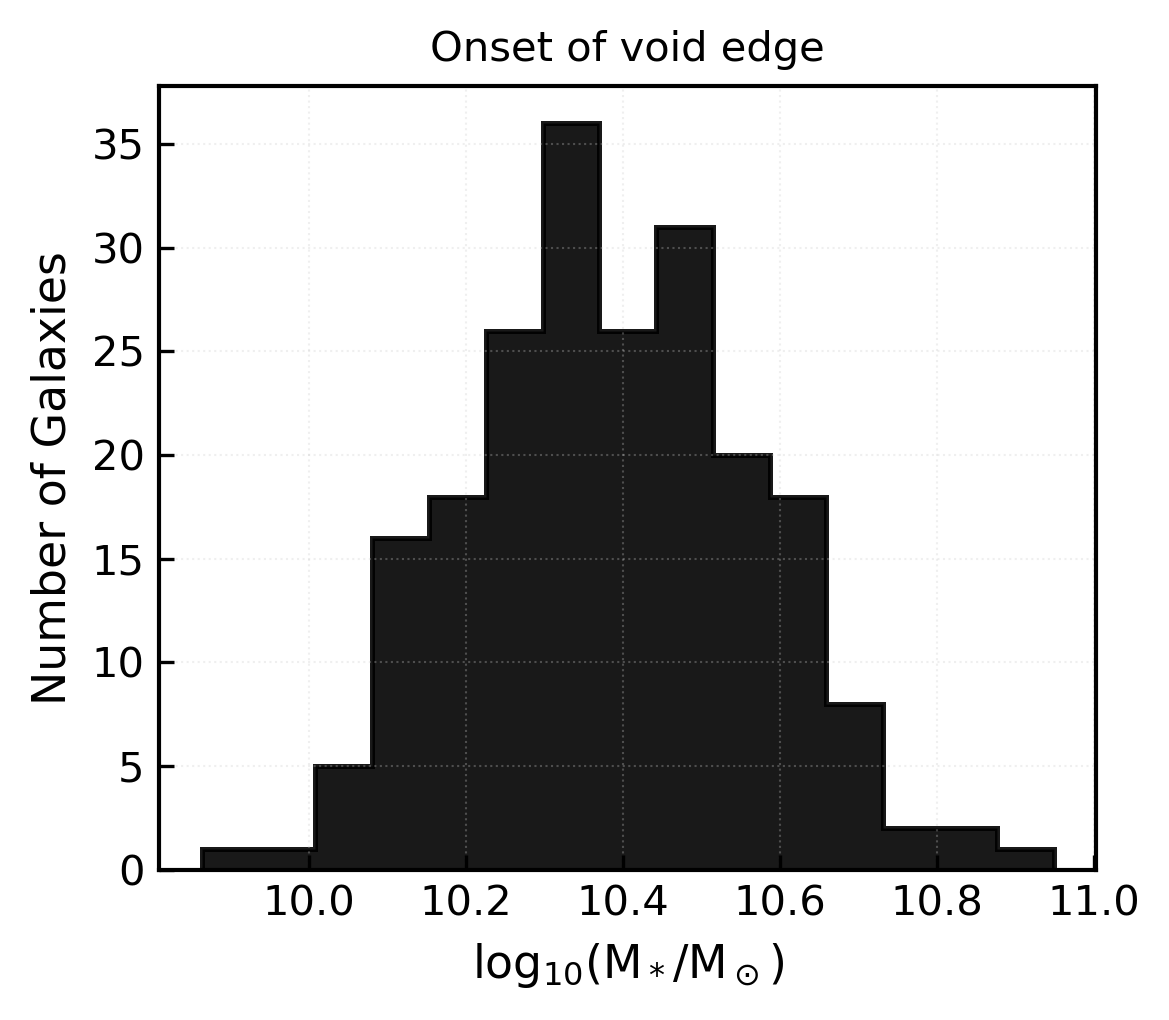} &
\includegraphics[width=0.31\textwidth]{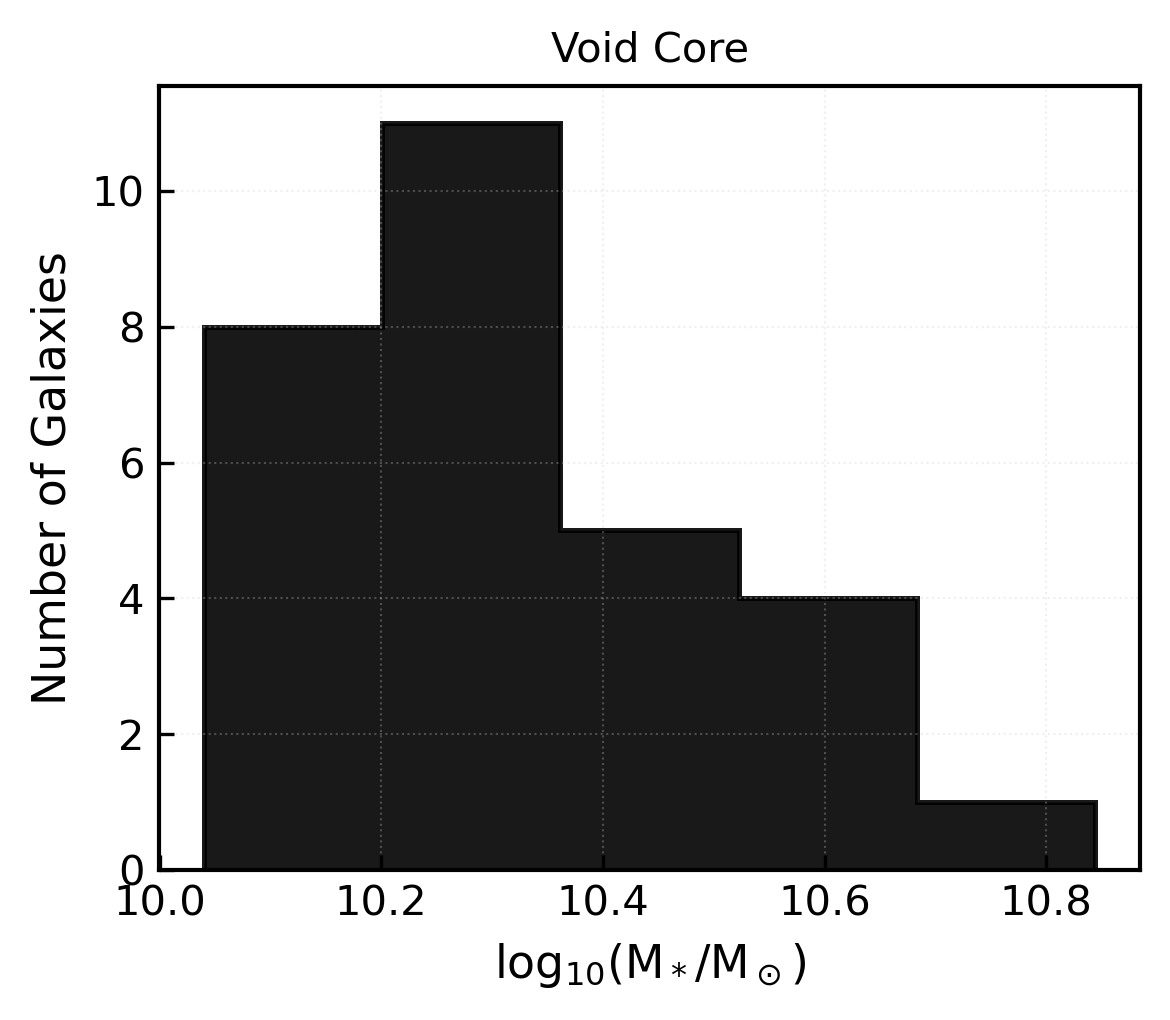} \\

\end{tabular}
\caption{Stellar mass distribution of void galaxies at different radial positions}
\label{fig:stellar_mass}
\end{figure*}

We observe a subtle but systematic increase in stellar mass towards the void edge shown in Figure~\ref{fig:stellar_mass}. The mean  rises from $10.33$ in the void core to $10.41$ in the outskirts , a difference of $0.08$ dex corresponding to an approximately $20\%$ increase in mass. The median masses follow the same trend, increasing from $10.31$ to $10.39$.

\subsection{Optical Color Properties}
\label{ssec:colors}

\begin{figure*}[hbt!]
\centering
\begin{tabular}{ccc}
\includegraphics[width=0.31\textwidth]{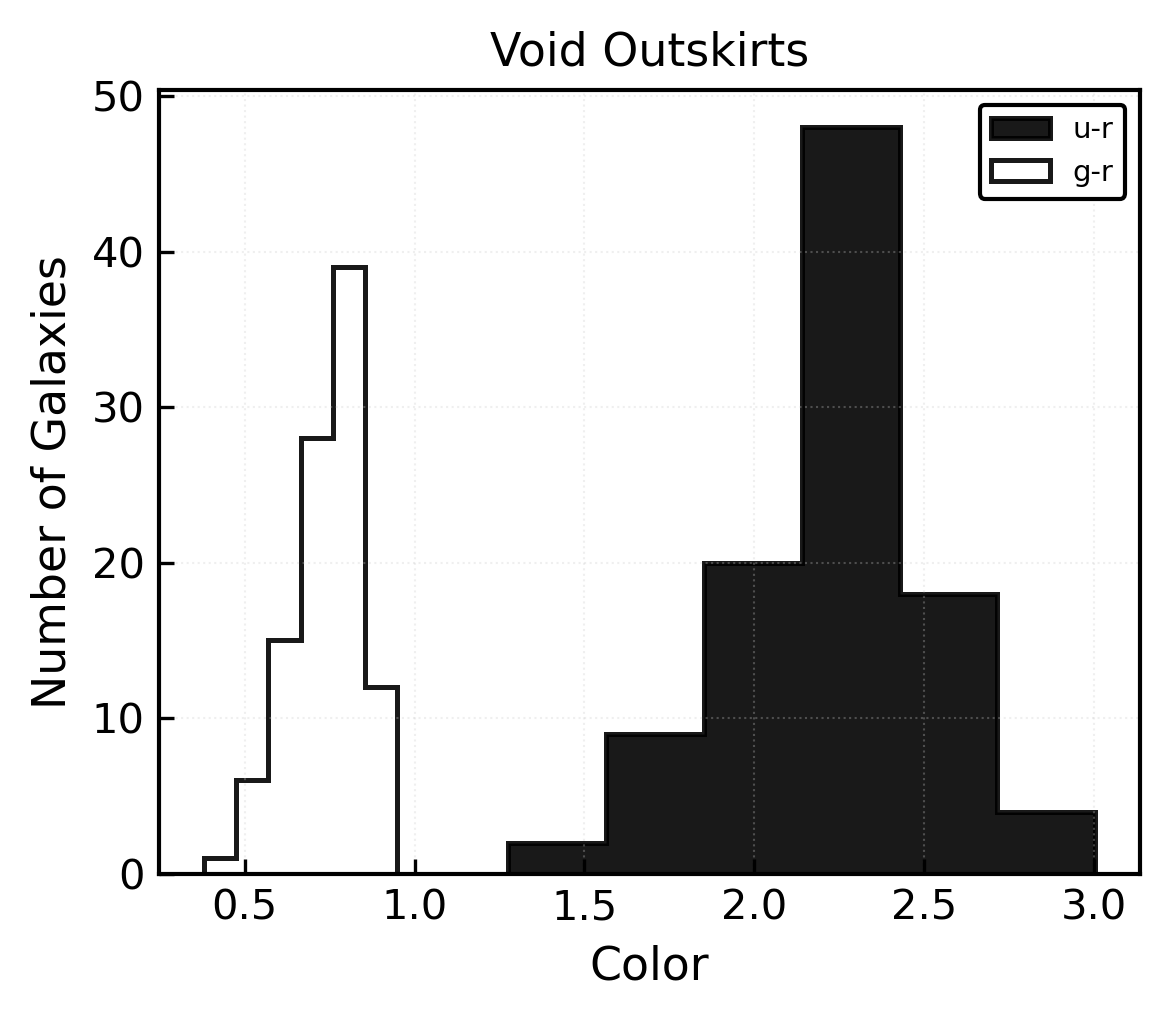} &
\includegraphics[width=0.31\textwidth]{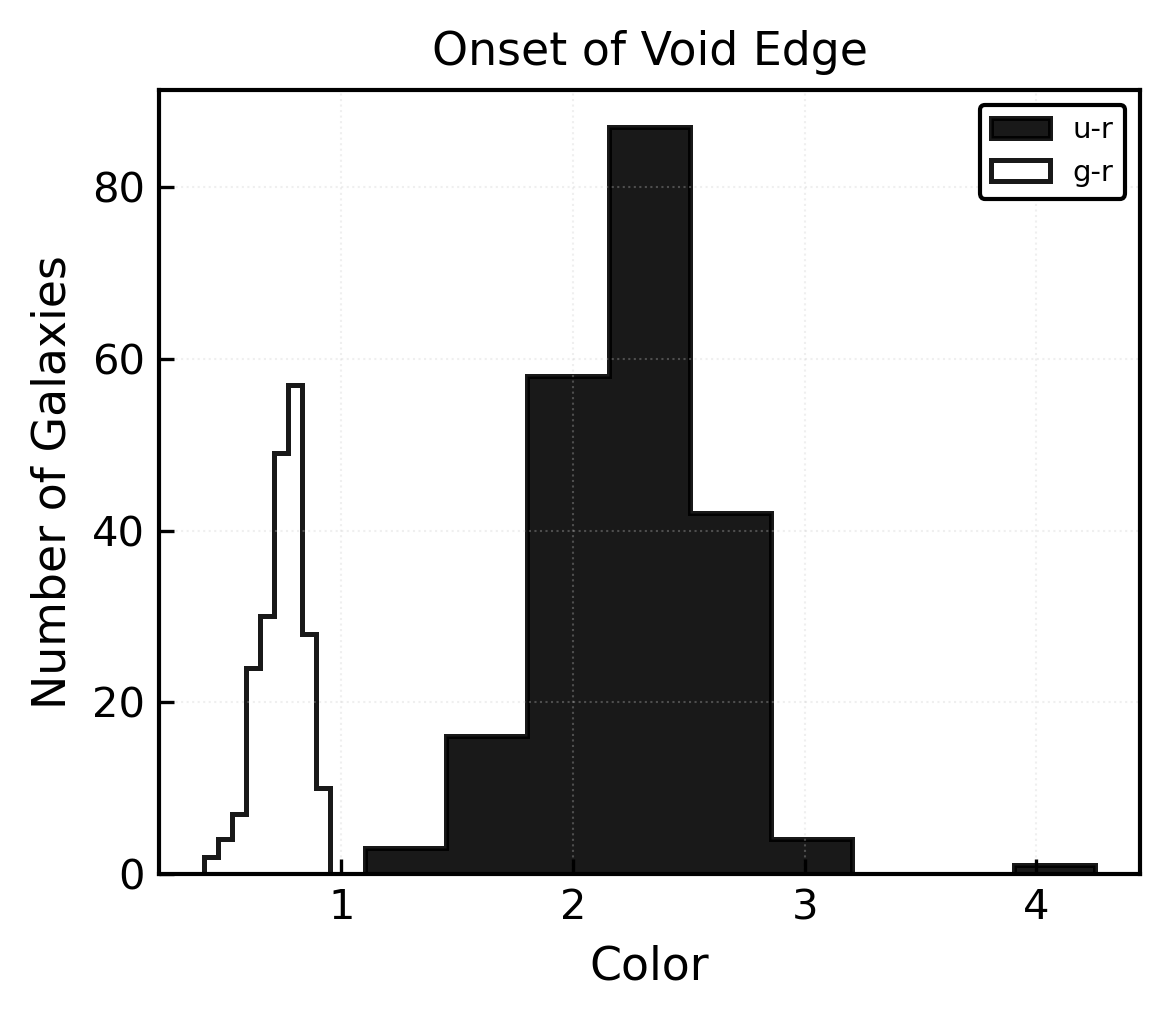} &
\includegraphics[width=0.31\textwidth]{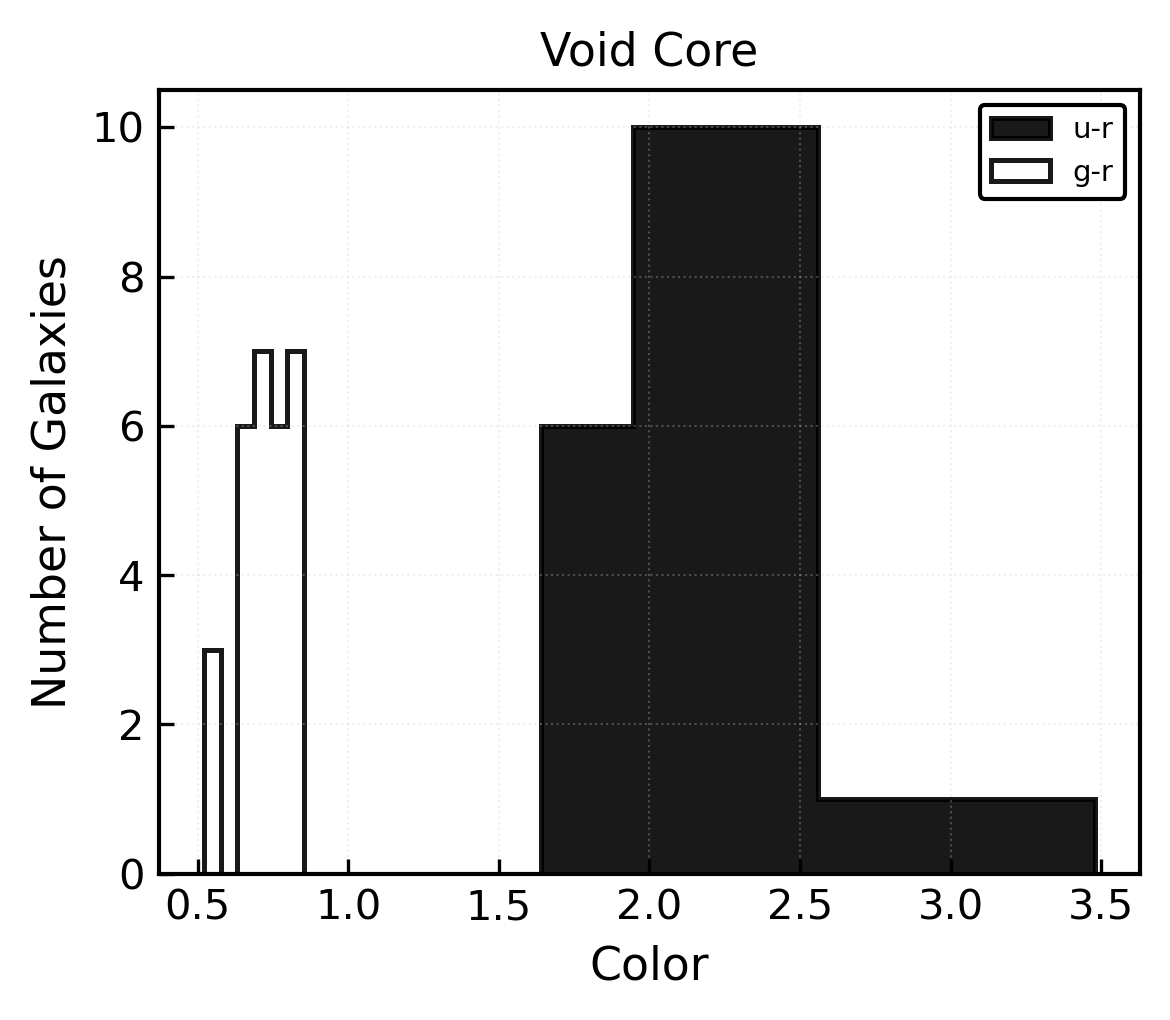} \\

\end{tabular}
\caption{Color properties of void ring galaxies.}
\label{fig:colors}
\end{figure*}

Figure~\ref{fig:colors} presents the color properties of our void ring galaxy sample. We find no significant radial variation in either $g-r$ or $u-r$ color. The median colors remain nearly constant across all bins, with $g-r$ around $0.7$–$0.8$ and $u-r$ around $2.2$–$2.3$. The blue fractions, defined by $g-r < 0.7$ and $u-r < 2.3$, also show no systematic trend with void radius, staying at approximately 30\% and 55\% respectively (Table~\ref{tab:physical_properties}). The void center does not host a distinctly bluer population; any apparent differences are well within the uncertainties.

\subsection{Star Formation Rates and Specific Star Formation Rates}
\label{ssec:sfr}

To quantify the star formation activity in our sample, we analyse the star formation rates (SFR) and specific star formation rates (sSFR). The sSFR, SFR per unit stellar mass, measures the efficiency of star formation relative to the already accumulated stellar mass.

Neither SFR nor sSFR shows a strong radial dependence (see Table~\ref{tab:physical_properties}). The median values hint at a slight decline toward the edge, but the large scatter and non‑monotonic behaviour of the means make any trend statistically inconclusive. More importantly, the fractions of star‑forming, passive, and green‑valley galaxies remain nearly constant across all radial bins. The passive fraction consistently dominates at $\sim 65\%$, while the star‑forming fraction stays below $6\%$ everywhere. 

\begin{figure*}[hbt!]
\centering
\begin{tabular}{ccc}
\includegraphics[width=0.31\textwidth]{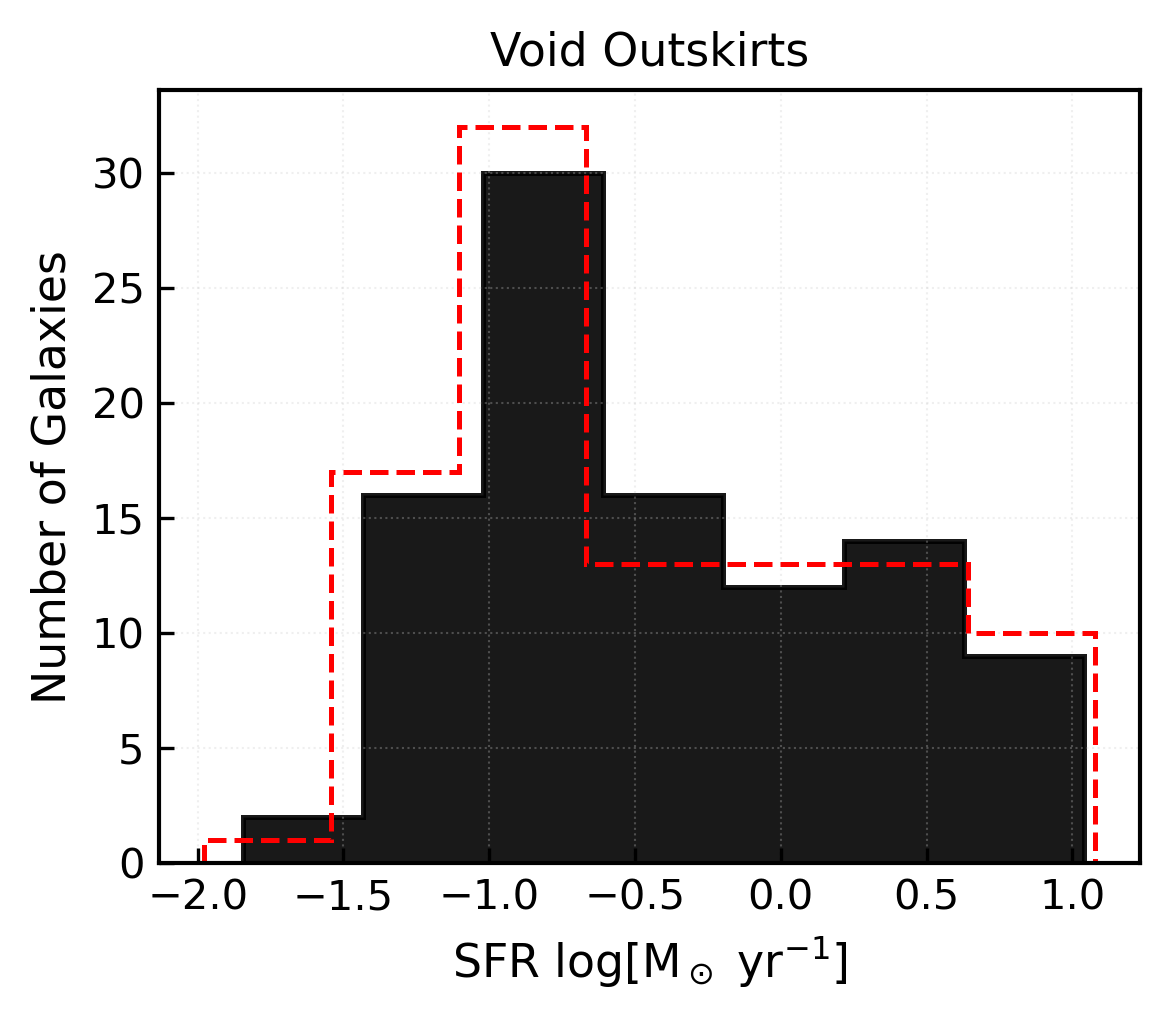} &
\includegraphics[width=0.31\textwidth]{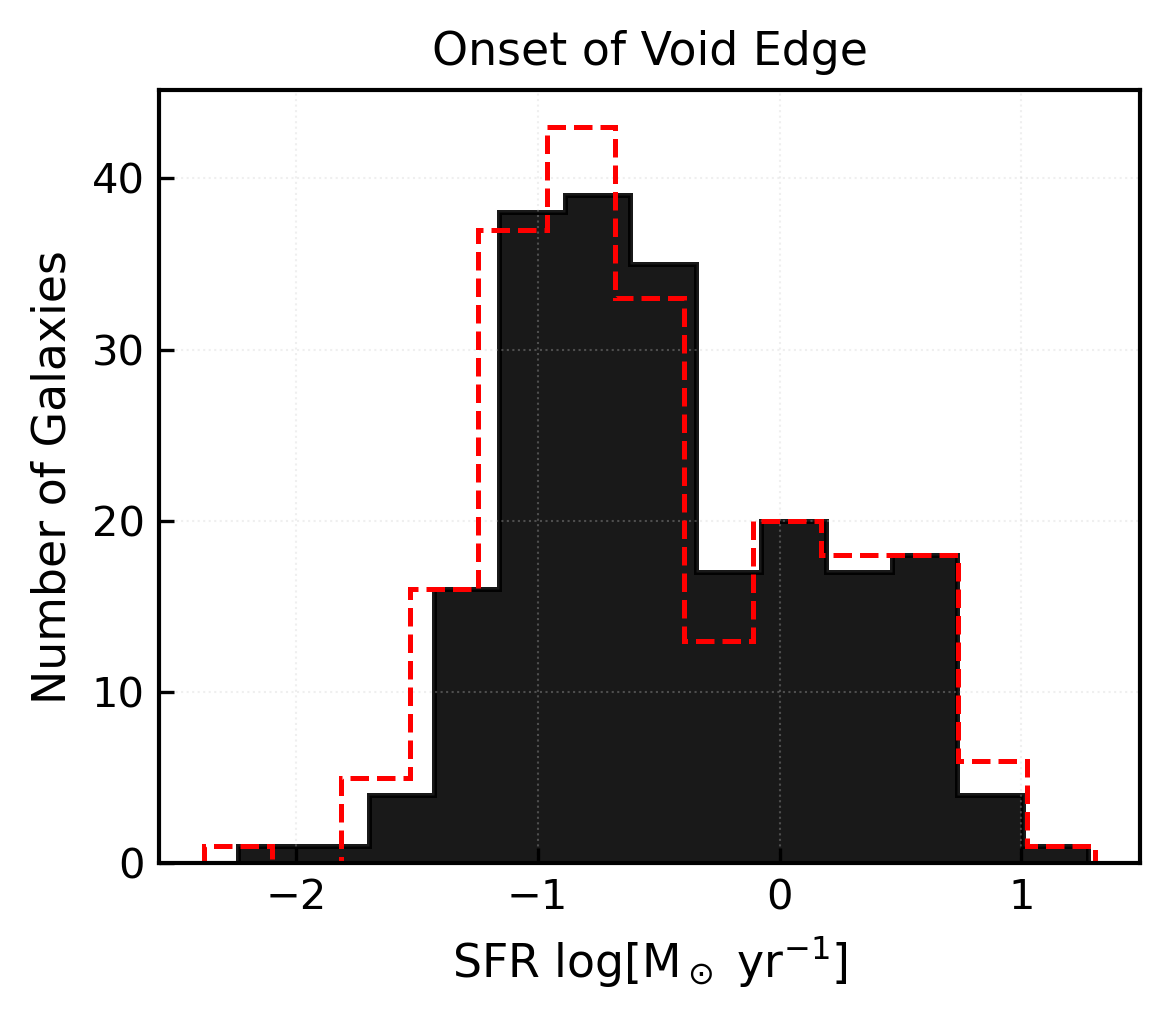} &
\includegraphics[width=0.31\textwidth]{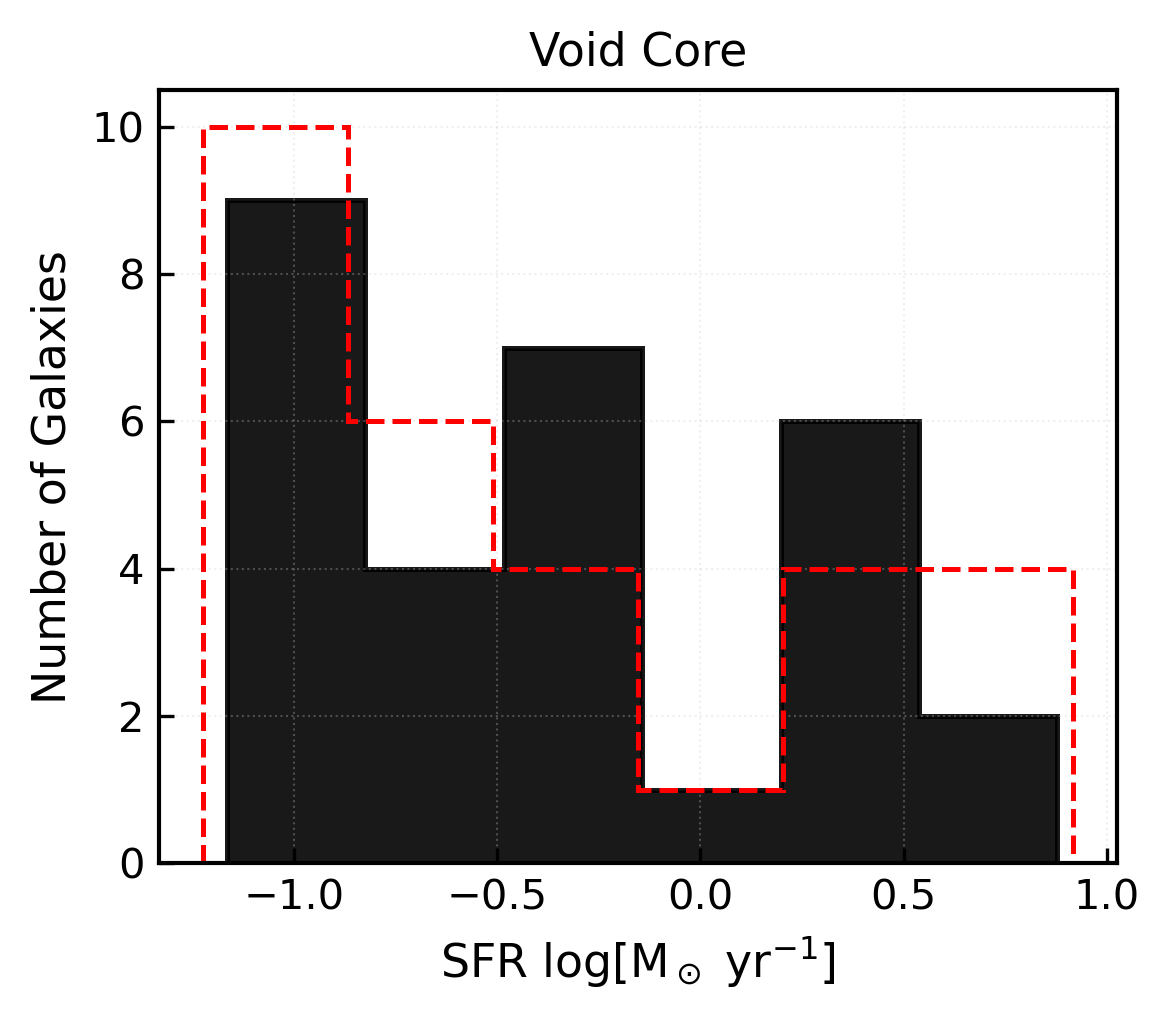} \\

\end{tabular}
\caption{SFR of void ring galaxies. The dotted histogram is of average SFR values and black filled histogram is of median SFR values.}
\label{fig:sfr}
\end{figure*}

\begin{figure*}[hbt!]
\centering
\begin{tabular}{ccc}
\includegraphics[width=0.31\textwidth]{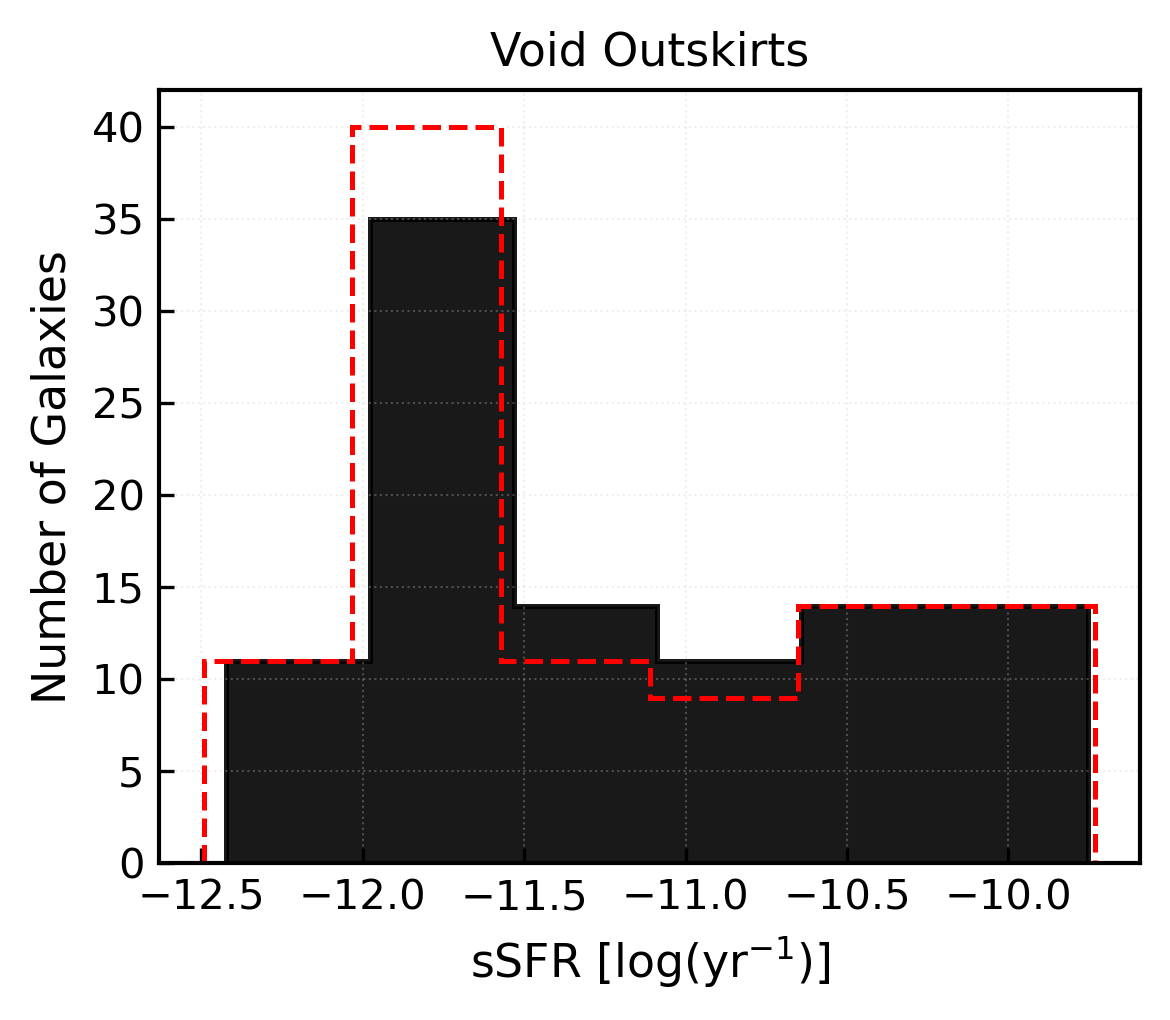} &
\includegraphics[width=0.31\textwidth]{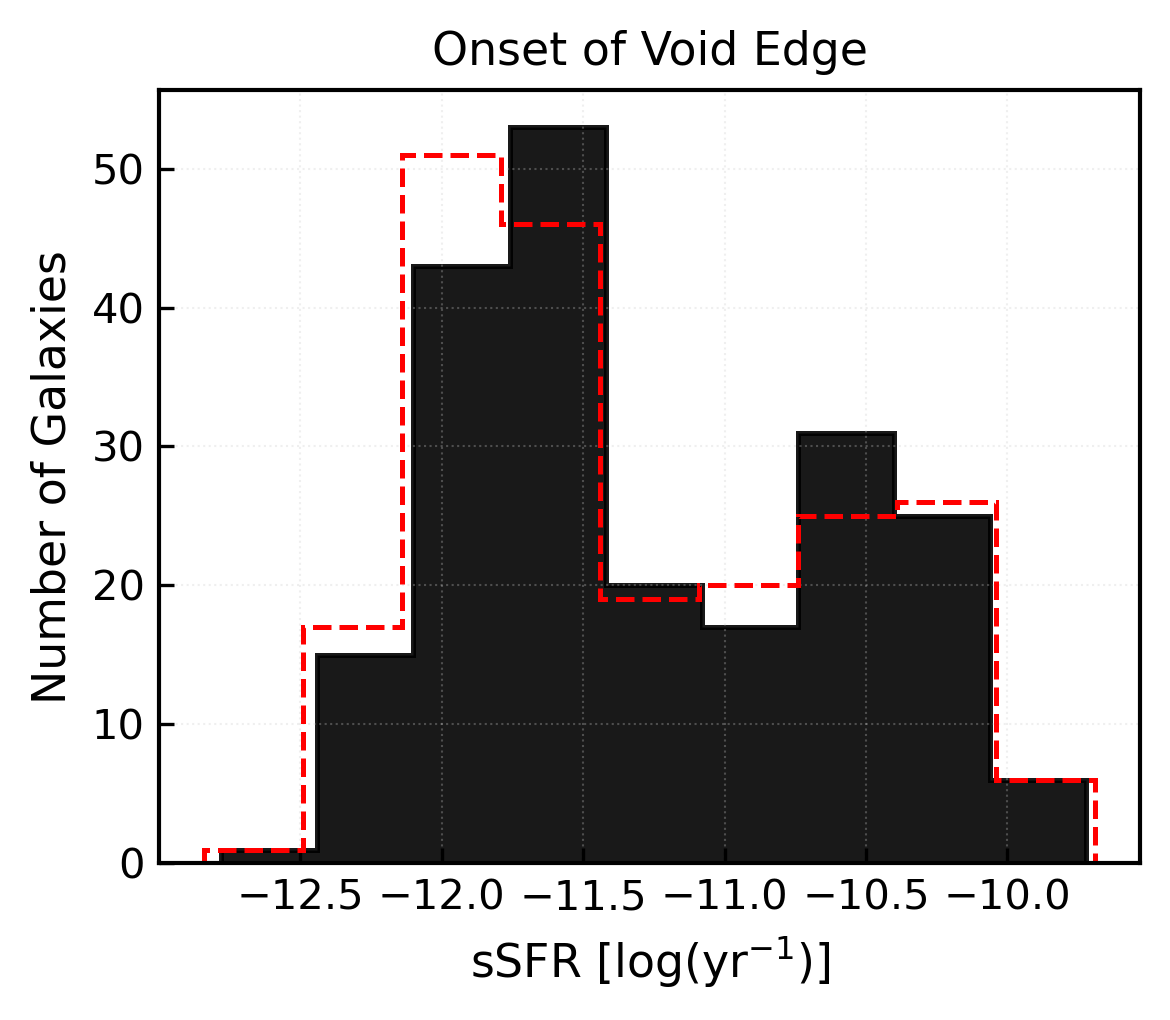} &
\includegraphics[width=0.31\textwidth]{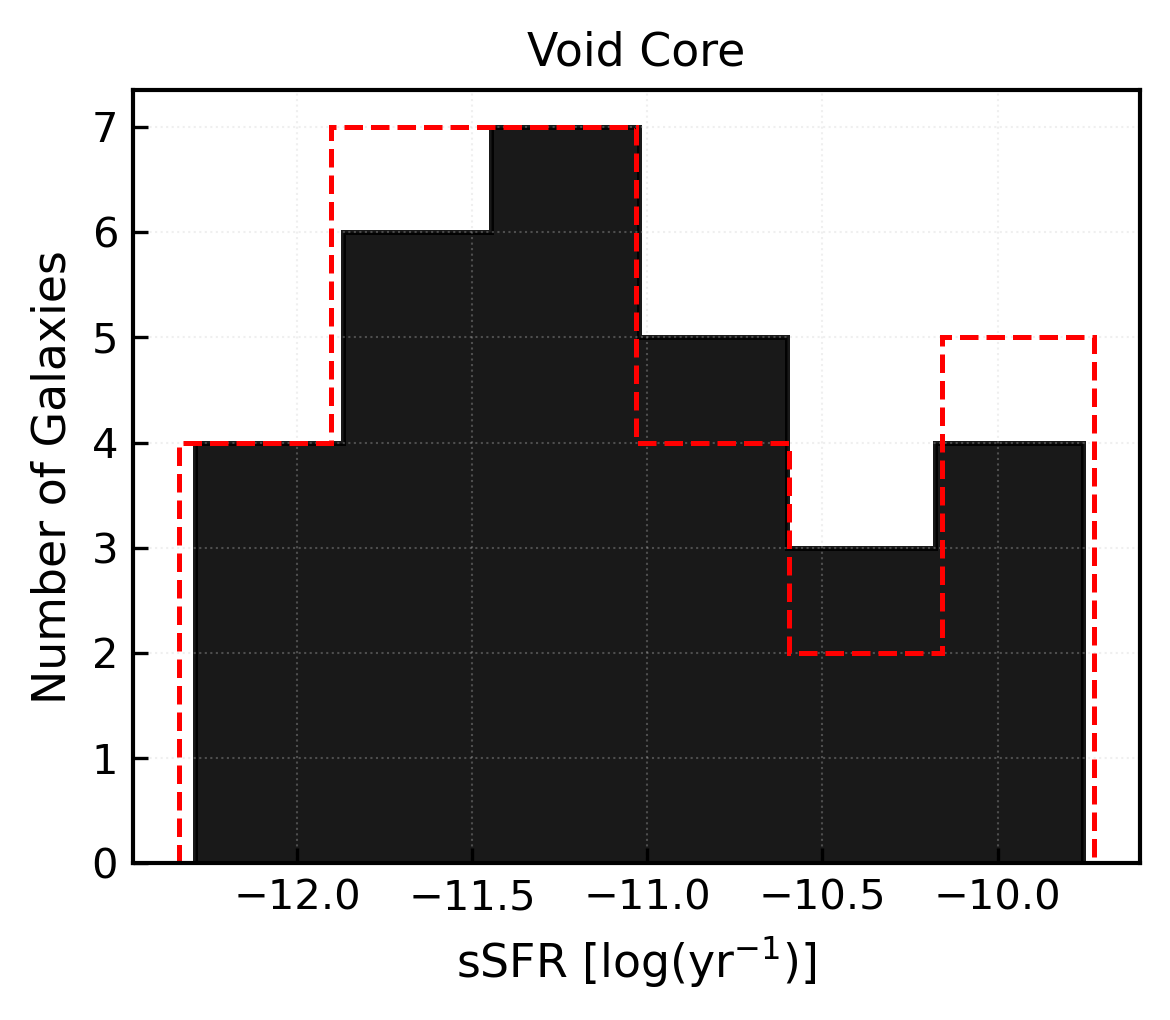} \\

\end{tabular}
\caption{sSFR of void ring galaxies.The dotted histogram is of average sSFR values and black filled histogram is of median sSFR values.}
\label{fig:ssfr}
\end{figure*}

\subsection{Comparson of properties with average void galaxies}
\label{comparison}

On comparison to general void population we observe that the void ring galaxies are more massive than the typical void galaxy. The mean stellar mass of our full sample is approximately $+0.6$ dex higher than the average for void galaxies. They are also significantly redder in optical color, with a mean $u - r$ color $\sim 0.5$ mag redder than the general void population.

Furthermore, the star formation activity of ring galaxies appears more subdued. Their mean specific star formation rate (sSFR) is about $-0.76$ dex lower than that of average void galaxies. This indicates that, while residing in the same underdense environments, ring galaxies represent a distinct subset that is more evolved, more massive, and less actively star-forming than the typical blue, faint, and high sSFR galaxies that dominate the void population.

\section{Discussions} \label{sec:highlight}
Our analysis of ring galaxies within cosmic voids presented us with a population whose morphology, distribution, and physical properties are sculpted by the most underdense environment in the universe. Below, we interpret our results in the context of prevailing theories of ring formation and environmental galaxy evolution.

\subsection{The Radial Imbalance}
\label{subsec:radial_imbalance}
The  radial distribution of ring galaxies within voids, with only 8.5\% in the core and over 91\% in the outer regions is the notable environmental signal in our study. This shows that the void edge represents a favourable spot for the existence of ringed structures. We discuss two possible interpretations for this imbalance:

\begin{enumerate}
\renewcommand{\labelenumi}{(\roman{enumi})}
    
    \item {Formation Efficiency:} The deep void interior i.e. void core may be too isolated. While resonant secular processes driven by internal bars can occur anywhere, the probability of even a minor interaction, a weak tidal encounter or the accretion of a tiny satellite  which can also trigger or amplify ring structures\citep{1996FCPh...16..111A,mapelli2012ring}, seems very unlikely in the void center. The edge region, while still underdense compared to walls, resides closer to the nascent density fluctuations that define the void boundary. Here, the frequency of very weak, non-destructive interactions may be just sufficient to instigate ring formation in susceptible disks, without being so frequent as to destroy the delicate structures afterward.

    \item {Survival and Detection:} Alternatively, ring galaxies may form at all radii, but their longevity could be environment-dependent. The pristine isolation of the void center might allow rings to persist for longer timescales, fading slowly due to secular diffusion. However, dynamical timescales are also longer in the lower-density halo potentials of the smallest void-center galaxies, potentially slowing the very internal evolution that creates and maintains rings. The edge, with its slightly higher density of dark matter and perhaps a gentler large-scale tidal field from the void's underdensity itself, might provide conditions more conducive to both the assembly of more massive disks as observed and the sustained dynamical activity needed to maintain a visible ring structure over Gyr timescales.
\end{enumerate}

This distribution aligns with the broader morphology-density relation, but in reverse: the denser region within the void mainly its its edge hosts a higher number density of our morphological type. It suggests that ring galaxies, often considered products of isolation, actually favor a threshold level of minimal environmental influence over absolute solitude.

\subsection{Morphological Characteristics}
\label{subsec:morphology}

The morphological breakdown gives us the look on the dominant evolutionary mechanism in voids. We see that inner rings and inner pseudorings together account for 45.2\% of all void ring galaxies (35.2\% and 10\%, respectively).Also Pure spiral structures without inner rings comprise only 8.8\% of the sample indicating that most ring galaxies in voids have developed inner ring features. Notably, inner lenses are also well-represented at 27.3\% with a consistent presence across all radial bin.

We see outer ring features are also widespread. Closed outer rings appear in  17.9\% of galaxies, while outer pseudorings are found in 56.3\% of galaxies. The presence of OLR subclass rings in galaxies tells us  that some systems have evolved to the point where outer Lindblad resonances have shaped their outer disks.

Moreover, nuclear features are rare in our sample, suggesting that that bar-driven gas inflow has not reached the central regions of many void ring galaxies or that such features are intrinsically uncommon in this sample

In the literature, inner rings are tightly coupled to the dynamics of galactic bars and are considered characteristic features of secular evolution\citep{1981ApJ...247...77S}. They form at internal resonance points (e.g., the Inner Lindblad Resonance) where bar-driven gas flows pile up, triggering star formation. Our results indicates that in the cosmic void environment, where major mergers are virtually absent and even intermediate mergers are exceedingly rare, this slow, internal process is the principal architect of galactic rings. The consistency of the prevalence of both the inner rings and outer rings  across all radial bins implies that when a void galaxy does develop a ring structure, it is almost due to its own internal dynamics, not an external trigger.

\subsection{Physical Gradients}
\label{subsec:gradients}

The subtle but systematic trends in stellar mass, color, and star formation rate from the void center to its edge is showing how  the weakest environmental gradient can influence galaxy evolution. It is important to note that even at the void center, our ring galaxies are more massive and redder than the average void galaxy (see Table~\ref{tab:physical_properties}), underscoring that ring structures are associated with a more evolved galactic substrate.
The increase in mean stellar mass (\(\Delta \log M_* \approx +0.08\) dex) towards the void edge is consistent with the universal trend of larger galaxies residing in deeper potential wells\citep{kelvin2014galaxy}, which are more likely to be found nearer the void boundary. This mirrors, on a much gentler scale, the well-known mass-density relation seen in clusters\citep{peng2010mass}. Galaxies at the void edge likely formed in slightly denser initial fluctuations and have had marginally greater access to accretion flows along tenuous filaments, leading to more massive disks.
 The complementary trends, a very weak decline in the blue fraction and a slight decrease in  specific SFR from center to edge suggest a very slow, density-dependent transition towards quiescence. The void center galaxies, in their extreme isolation, appear to be the most delayed in their evolution. They retain a slightly higher specific star formation rate, consistent with the downsizing paradigm where lower-mass systems in lower-density environments form their stars over more extended \citep{avila2007understanding}. The edge galaxies, while still gas rich compared to cluster populations, have progressed further along the path of gentle gas consumption or stabilization, leading to redder colors and lower sSFRs. We see that this is not the rapid quenching but a gentle fading induced by a combination of slightly more efficient past star formation building mass and perhaps a marginally greater susceptibility to very weak gas-stripping or starvation processes at the void boundary.

\section{Conclusions} \label{sec:cite}
We studied the morphological and physical properties of ring galaxies residing inside cosmic voids. Our analysis of 341 void ring galaxies shows  that ring structures are not randomly distributed within these underdense regions. Instead, we find a strong radial preference, the majority are located near the void edges, while the deep interior hosts only few.

In regard to morphological composition of our sample, we find that inner rings and inner pseudorings are prominent (45.2\% combined ) of all void ring galaxies, while outer pseudorings rings are even more common and closed outer rings appear in 17.9\% of galaxies. Nuclear features are rare and we see multiple outer features more frequent toward void centers. This morphological signature indicates that secular evolution, driven by internal dynamics such as bars and resonances, is the primary mechanism for ring formation in these tranquil environments, rather than interactions or mergers.

Furthermore, we also detected subtle environmental gradients i.e. stellar mass increases by \(\Delta\log M_* \approx +0.08\) dex toward the void edge, while specific star formation rates show a gentle decline. When compared to the general void galaxy population, our ring galaxies are markedly more massive, redder, and have lower specific star formation rates which shows  them as a distinct, more quiescent subsample within voids. We conclude that these trends showed a gentle, density-dependent evolutionary progression within voids. Our results shows void ring galaxies as a population defined by secular processes and highlight how even the weakest environmental gradient can systematically shape galactic properties.

\begin{acknowledgments}

We are grateful to the Sloan Digital Sky Survey and NASA-Sloan Atlas (NSA) for providing public access of their spectroscopic and photometric data. We also thank the Galaxy Zoo team and citizen scientists around the world whose morphological classifications have been used in this work. The void catalog used in this analysis was generated by the VoidFinder algorithm applied to the SDSS data, and we thank the authors for making it publicly available. We acknowledge the use of the AASTeX template for preparing this manuscript.
\end{acknowledgments}

\software{Astropy \citep{price2022astropy},  
          Numpy \citep{harris2020array}, 
          Scikit-learn \citep{scikit-learn},
         Matplotlib \citep{Hunter:2007},
        TOPCAT \citep{2005ASPC..347...29T} }

\bibliography{paper701}{}
\bibliographystyle{aasjournalv7}

\end{document}